\documentclass[12pt]{iopart}
\expandafter\let\csname equation*\endcsname\relax
\expandafter\let\csname endequation*\endcsname\relax
\usepackage{amsmath}
\usepackage{graphicx} 
\usepackage{layout}
\usepackage{caption}
\usepackage{pgfplots}
\pgfplotsset{compat=1.3}
\usepackage[hidelinks, bookmarks=True]{hyperref}
\hypersetup{colorlinks,linkcolor=blue}
\usepackage{amsmath}
\usepackage{amssymb}
\usepackage{siunitx}
\usepackage{commath}
\usepackage{mathtools}
\usepackage{color}
\usepackage{listings}
\usepackage[english]{babel}
\usepackage[autostyle, english = american]{csquotes}
\MakeOuterQuote{"}
\usepackage{enumitem}
\usepackage{amsmath,mleftright}
\usepackage{xparse}
\usepackage{tabularx}
\usepackage{babel}
\usepackage{subcaption}
\usepackage{cleveref}
\usepackage{graphicx}
\usepackage{pdfpages}
\usepackage{multirow}
\usepackage{multicol}
\usepackage{xcolor}
\usepackage{multicol}
\usepackage{breqn}
\usepackage{afterpage}
\usepackage{bookmark}

\usepackage{tikz}
\definecolor{lime}{HTML}{A6CE39}
\DeclareRobustCommand{\orcidicon}{
    \hspace{-3mm}
	\begin{tikzpicture}
	\draw[lime, fill=lime] (0,0) 
	circle [radius=0.16] 
	node[white] {{\fontfamily{qag}\selectfont \tiny ID}};
	\draw[white, fill=white] (-0.0625,0.095) 
	circle [radius=0.007];
	\end{tikzpicture}
	\hspace{-4mm}
}

\foreach \x in {A, ..., Z}{%
	\expandafter\xdef\csname orcid\x\endcsname{\noexpand\href{https://orcid.org/\csname orcidauthor\x\endcsname}{\noexpand\orcidicon}}
}

\definecolor{javared}{rgb}{0.6,0,0} 
\definecolor{javagreen}{rgb}{0.25,0.5,0.35} 
\definecolor{javapurple}{rgb}{0.5,0,0.35} 
\definecolor{javadocblue}{rgb}{0.25,0.35,0.75} 
 
\begin{document}
\title[]{CNN-Based Vortex Detection in Atomic 2D Bose Gases in the Presence of a Phononic Background}

\author{Magnus Sesodia\orcidA{}, Shinichi Sunami\orcidB{}, En Chang\orcidC{}, Erik Rydow\orcidD{}, Christopher J. Foot\orcidE{} and \mbox{Abel Beregi\orcidF{}}}
\address{Clarendon Laboratory, University of Oxford, Parks Road, Oxford OX1 3PU, United Kingdom}
\ead{abel.beregi@physics.ox.ac.uk}
\vspace{10pt}

\begin{abstract}
Quantum vortices play a crucial role in both equilibrium and dynamical phenomena in two-dimensional (2D) superfluid systems.
Experimental detection of these excitations in 2D ultracold atomic gases typically involves examining density depletions in absorption images, however the presence of a significant phononic background renders the problem challenging, beyond the capability of simple algorithms or the human eye.
Here, we utilize a convolutional neural network (CNN) to detect vortices in the presence of strong long- and intermediate-length scale density modulations in finite-temperature 2D Bose gases.
We train the model on datasets obtained from \textit{ab initio} Monte Carlo simulations using the classical-field method for density and phase fluctuations, and Gross-Pitaevskii simulation of realistic expansion dynamics.
We use the model to analyze experimental images and benchmark its performance by comparing the results to the matter-wave interferometric detection of vortices, confirming the observed scaling of vortex density across the Berezinskii-Kosterlitz-Thouless (BKT) critical point.
The combination of a relevant simulation pipeline with machine-learning methods is a key development towards the comprehensive understanding of complex vortex-phonon dynamics in out-of-equilibrium 2D quantum systems.
\end{abstract}

\maketitle

\section{Introduction}

Quantum vortices are topological excitations found in superfluids and are characterized by integer multiples of $\pm 2\pi$ phase winding around a vortex core where the density vanishes.
They have been experimentally observed in a wide variety of systems including liquid helium \cite{Vinen_1961, Tang_2023}, type-II superconductors \cite{Essmann_1967, tonomura_2001, roditchev_2015, Llorens_2020}, photon fields \cite{Genevet_2010}, exciton-polariton superfluids \cite{Lagoudakis_2008, Choi_2022}, and in ultracold atom systems with bosonic \cite{Matthews_1999, Stock_2005, Hadzibabic_2006} and fermionic \cite{Zwierlein_2005, liu_2021, lee_2024} particles.
The precise spatial and temporal control of the ultracold atom platform facilitates the study of numerous manifestations of vortex excitations as a consequence of rich static and dynamical many-body phenomena -- for example, vortices have been shown to arrange into lattices in rotating Bose-Einstein condensates (BECs) \cite{Abo-Shaeer_2001} and randomly form following a rapid quench into an ordered phase \cite{liu_2021, lee_2024, Weiler_2008, Chomaz_2015, Rabga_2023}.
An important class of vortex excitations emerges in systems within the 2D XY universality class to which ultracold 2D Bose gases belong \cite{Posazhennikova_2006}, having characteristic paired and unpaired configurations across the superfluid transition \cite{Berezinskii_1972, Kosterlitz_1973} (see \mbox{\ref{sec:bkt_transition}} for a brief overview).
The statistics and spatial structure of vortex excitations are one of the direct measures of the many-body states and their time-evolution, hence the precise determination of the number and location of vortices is critical for studying superfluid dynamics.
Absorption imaging is typically used to obtain images of the density of ultracold atomic systems \cite{Ketterle_1999, Hueck_2017}, however the task of identifying vortices from these is often non-trivial due to their small spatial structure and misleading patterns arising from other excitations in the system.
In some situations, it can be simple to count vortices by eye, e.g. in the ground state of 3D rotating BECs where these vortices form an ordered triangular lattice \cite{Abo-Shaeer_2001, Engels_2004, Williams_2010}.
However, in stationary states of finite-temperature 2D Bose gases the vortices do not form a regular pattern and the presence of phonons, which can result in similar density modulations after time-of-flight expansion \cite{Choi_2012, Sunami_2024}, makes vortex detection by the human eye very inaccurate and impossible to use for quantitative analysis.
Matter-wave interferometry addresses this challenge by taking advantage of the characteristic $\pm \pi$ phase discontinuity across vortices (singly quantized) \cite{Stock_2005, Hadzibabic_2006, Sunami_2022}.
These distinct features are easy to distinguish from other excitations which have a significantly longer characteristic length scale. 
This approach has accelerated 2D Bose gas research in recent years by providing a method to obtain the average vortex density from an ensemble of images \cite{Sunami_2022, Sunami_2023, Beregi_2024}, however this method is limited to measuring averages, missing the rich information available in single-shot detection of quantum vortex locations distributed in the plane of the 2D gas.

To address this constraint, here, we report on the development of a robust method for detecting vortex excitations in atomic 2D Bose gases by employing machine-learning techniques.
In particular, we focus on the experimentally-relevant regime where thermally-activated vortices exist intermingled amongst other intrinsic excitations such as phonons which renders vortex detection by the human eye or simple algorithms challenging.
Metz \textit{et al.} \cite{Metz_2021} showed that a convolutional neural network (CNN) trained on synthetic data of rotating 3D BECs can detect vortices in out-of-sample data, even in the presence of technical noise.
Recently, Kim \textit{et al.} \cite{Kim_2023} applied the same approach to turbulent condensates, and accurately detected vortices in experimental data.
Here, we focus on the challenging situation of detecting vortices arising from the BKT phase transition, which are obscured by the presence of long- and intermediate-length scale density fluctuations by deploying a more comprehensive training and inference pipeline that is robust to both technical and intrinsic noise.
We report automated vortex detection in experimental images well beyond the capability of counting by eye, and benchmark against the results obtained via matter-wave interferometry \cite{Sunami_2022}.
The investigation of phase transitions at finite temperature is an important application for which our method can identify vortices in a non-uniform background density.
More generally, we establish a framework for combining simulation with machine-learning methods which can be applied to a wide variety of systems.

The remainder of this paper is organized as follows.
Section \ref{sec:methods} describes how we numerically simulate 2D Bose gases to obtain experimentally-relevant training data.
Section \ref{sec:cnn_design} outlines the design and operation of our CNN.
Section \ref{sec:results} evaluates the model's performance against validation data and examines its effectiveness at predicting vortex locations in experimental images.
Finally, section \ref{sec:conclusion} summarizes our work and provides an outlook for the present result.

\section{Synthesizing training data}\label{sec:methods}

\subsection{Numerical simulation of 2D Bose gases}

The physics of ultracold 2D Bose gases can be described by the Hamiltonian 

\begin{equation}\label{eq:2d_bose_gas_quantum_hamiltonian}
    \hat{H} = \int d\textbf{r} \left[\frac{\hbar^2}{2m} \boldsymbol{\nabla} \hat{\psi}^{\dagger}(\textbf{r}) \cdot \boldsymbol{\nabla} \hat{\psi}(\textbf{r}) + \frac{g}{2}\hat{\psi}^{\dagger}(\textbf{r})\hat{\psi}^{\dagger}(\textbf{r})\hat{\psi}(\textbf{r})\hat{\psi}(\textbf{r}) + V(\textbf{r})\hat{\psi}^{\dagger}(\textbf{r})\hat{\psi}(\textbf{r}) \right],
\end{equation}

\noindent where $\hat{\psi}(\textbf{r})$ and $\hat{\psi}^{\dagger}(\textbf{r})$ are annihilation and creation operators for bosons, respectively, at position vector $\textbf{r}$ in the x-y plane, $g = \widetilde{g}\hbar^2/m$ is the quasi-2D interaction parameter, $\widetilde{g}$ is the dimensionless 2D interaction strength, and $V(\textbf{r})$ is the external potential \cite{Prokofev_2001}.
For the experimental parameters used the trapped gas exhibits near-homogeneous density, therefore, we considered a round box potential of radius 22.5 \textmu m.
This also allowed us to define the phase-space density of the gas as $\mathcal{D} = n h^2 /(2\pi m k_B T) \propto n/T$, where $n$ is the average 2D density, and $T$ is the temperature (see \mbox{\ref{sec:exp_details}} for more experimental details).

To generate \textit{in-situ} samples of the fluctuating Bose field for our system, we first used the classical field approximation to replace the field operators in (\ref{eq:2d_bose_gas_quantum_hamiltonian}) with complex numbers and their complex conjugates \cite{Blakie_2008}.
This approach accurately captured both vortex and phonon behavior, and has been previously used to support experimental observations \cite{Weimer_2015, Sunami_2022}. 
We mapped the system onto a $280 \times 280$ square lattice with lattice constant $\delta l =$ 0.25 \textmu m and evaluated the derivatives using first-order symmetric differences.
The lattice constant was chosen smaller than the characteristic length scales of the system to ensure accurate representation of the continuum limit (see \mbox{\ref{sec:simulation_verification}} for calculation). 

We employed the Metropolis-Hastings algorithm \cite{Metropolis_1953, Hastings_1970} to draw random states of a finite-temperature 2D Bose gas from the grand canonical ensemble.
We initialized the complex field with zero real and imaginary components.
At every iteration, we chose a random site and generated proposed updates to the real and imaginary parts of the on-site field by drawing two random numbers from a normal distribution of zero mean and $\sigma$ standard deviation.
We then calculated the grand-canonical weight \mbox{$p=\exp{ \left( -\beta \left[ \Delta E - \mu \Delta N \right] / k_BT \right)}$}, where $\mu$ is the chemical potential, $\Delta E,\Delta N$ are the changes in energy and number of particles in the system by the proposed update of the fields, and $k_B$ is the Boltzmann constant.
In accordance with the Metropolis-Hastings algorithm, the proposed field update was accepted if $p>1$ or if $p \leq x$, where $x$ was a random number drawn from a $\left[0,1 \right]$ uniform distribution.
Full details can be found in \ref{sec:simulation_verification}.

\begin{figure}[t]
    \centering
    \includegraphics[width=1.0\linewidth]{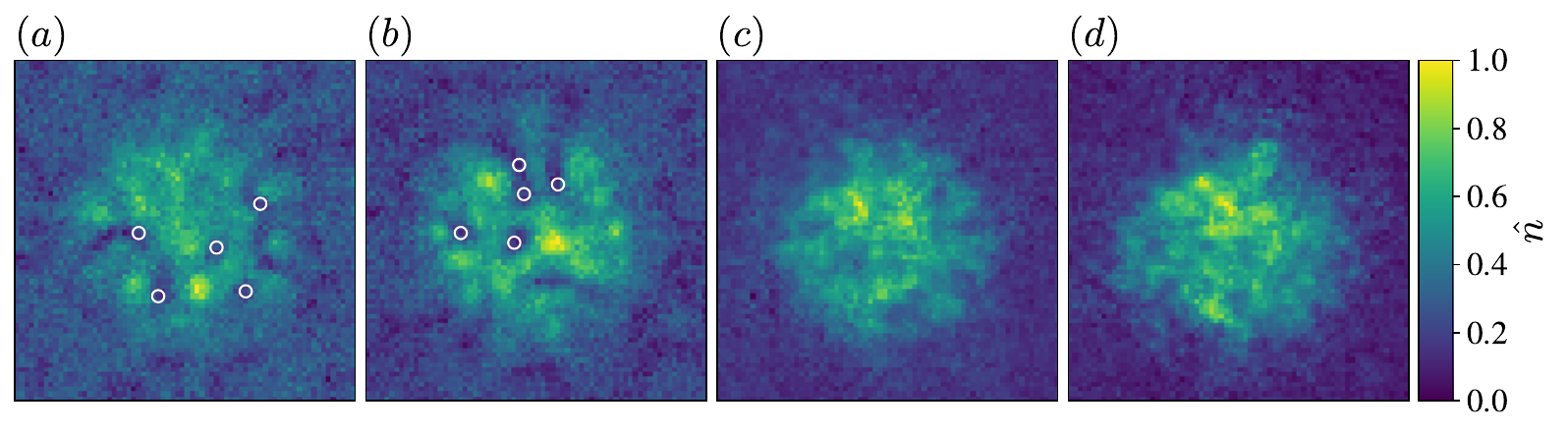}
    \caption{
    Comparison of density images: (a)-(b) training and (c)-(d) experimental. Our simulation captured the presence of phonons which manifest as density fluctuations and, in some cases, density dips with no associated vortices.
    The normalized density $\hat{n}$ ranges from 0 to 1 over each image.
    }
    \label{fig:train_vs_exp}
\end{figure}

\subsection{Data processing}

\begin{figure}[t]
    \centering
    \includegraphics[width=1.0\linewidth]{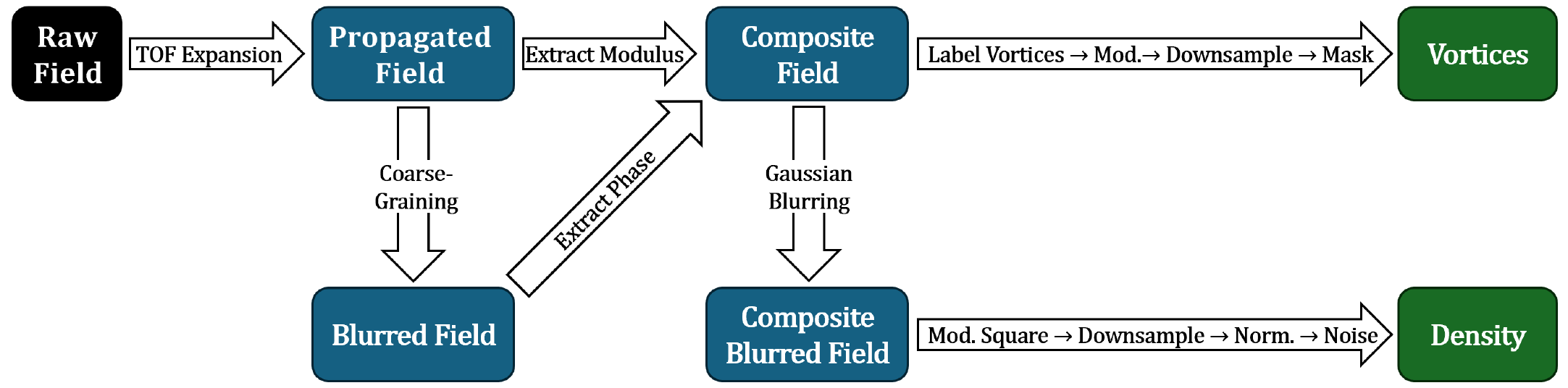}
    \caption{
    The processing steps performed on a single simulated field to obtain the density and vortex map for a single training example.
    }
    \label{fig:data_processing_pipeline}
\end{figure}

To prepare the output of the Monte Carlo (MC) simulations for training we applied the following steps, laid out in Figure \ref{fig:data_processing_pipeline}, to each field.
First, we simulated the short time-of-flight (TOF) expansion, as performed in the experiment, for 5.3\,ms which increased the size of the vortex cores. This involved numerically solving the time-dependent Gross-Pitaevskii equation (GPE) \cite{Dalfovo_1999, Brennecke_2019} using split-step Fourier methods \cite{Javanainen_2006}.
We performed simulations in 2D and ignored the interactions as they become negligible after the rapid expansion of the cloud in the z-direction which occurs in the experiment.
Simulating in 2D captured the important physics while being significantly less computationally demanding than 3D simulations; see \ref{sec:data_processing_methods} for further discussion.

Next, we blurred the field using a Gaussian filter of spatial width (standard deviation) $\sigma_1$ and constructed a composite field in the density-phase representation using the absolute value of the propagated (unblurred) field and the phase of the blurred field.
This coarse-graining \cite{Foster_2010} removed closely-bound vortex pairs and their associated density dips (length scales smaller than $\sigma_1$) which allowed us to detect the unbound vortices that are responsible for the superfluid transition, even as bound vortex-antivortex pairs proliferate.
From this composite field, we identified vortices by checking for the characteristic $\pm 2\pi$ phase-winding at each \mbox{$2 \times 2$} plaquette of sites which yielded a binary vortex map of shape \mbox{$280 \times 280$}.
We did not distinguish between vortices and antivortices since it is impossible to tell the direction of circulation solely from a density image.
Then, we downsampled the vortex map by a factor of four to match the pixel size of our experimental images and applied a mask to restrict vortices to within the central region of the density profile to capture the bulk properties of the system.
To emulate the finite resolution of the experimental imaging system, we applied a second Gaussian filter of spatial width $\sigma_2$ to the composite field and calculated the squared norm to obtain the density.
We applied the same factor-of-four downsampling as with the vortex map, but used the mean values instead of the maximum values of the corresponding \mbox{$4 \times 4$} plaquettes in the original image.
Then, we normalized density to the range [0, 1] and added Gaussian white noise of standard deviation $\sigma = 0.03$ to ensure our model was robust against the typical level of background noise present in experimental images.
Figure \ref{fig:train_vs_exp} shows two of the simulated images after undergoing processing in comparison with two experimental images.

\section{CNN Design}\label{sec:cnn_design}
 
\begin{figure}[t]
    \centering
    \includegraphics[width=1.0\linewidth]{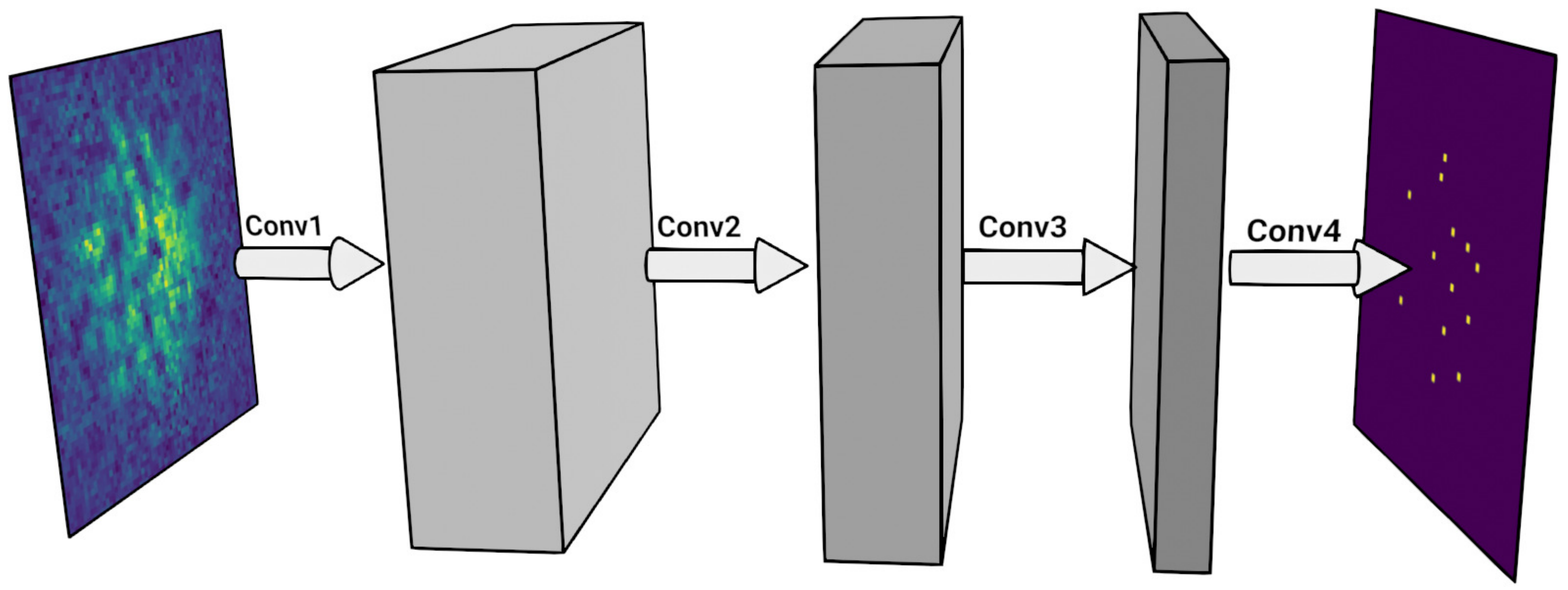}
    \caption{
    The neural network architecture consists of 4 convolutional layers with 30, 25, 20 and 1 filters, respectively, which transform density image inputs into vortex map outputs.
    }
    \label{fig:cnn_arch}
\end{figure}

We designed our CNN with the pixel size of the experimental images (around \mbox{1 \textmu m}) in mind.
Our CNN inputs a \mbox{70 $\times$ 70} grayscale image of density, with pixel size \mbox{1 {\textmu m}$^2$}, and passes it through four successive convolutional layers, outputting a vortex map, denoted $\widetilde{Y}$, which is a tensor of the same size where the value of each cell is the probability of a vortex being present in the equivalent cell of the input (see Figure \ref{fig:cnn_arch}).
The ground-truth, denoted $Y$, is also of the same size, but is restricted to the values 0 and 1.
The number of convolutional layers and filters was iteratively optimized to capture vortex patterns with high accuracy while mitigating overfitting to the training data.
This refinement ensured robust generalization to the experimental data while minimizing the computational overhead associated with over-parameterization.
Notably, omitting fully connected layers significantly reduced memory requirements, improving the model's efficiency and adaptability for deployment.

We used a weighted binary cross-entropy loss function which is commonly used in binary classification:

\begin{equation}\label{eq:loss}
    \mathcal{L}(\widetilde{Y}, Y) = \sum\limits_{\text{batch}} \sum\limits_{i,j}\left[- w_1 Y_{ij} \log{\widetilde{Y}_{ij}} - (1 - Y_{ij})\log{(1 - \widetilde{Y}_{ij})}\right] 
    ,
\end{equation}

\noindent where the sum $i,j$ is over the grid positions.
Since our training set contained significant class imbalance (on average, only 0.32\% of sites had vortices), we introduced the $w_1$ hyperparameter to encourage more positive predictions.
Without this, the model often fell into local minima of the loss function corresponding to predicting no vortices at all.
Further details of the CNN architecture and training choices can be found in \mbox{\ref{sec:cnn_details}}.

\section{Results}\label{sec:results}

\subsection{Validation}

Our validation set consisted of 1000 images (5\% of the training data) and their accompanying vortex locations.
To evaluate our model's predictions on these, we used precision, recall and F1 score which are standard metrics in binary classification and allowed for off-by-one errors where the prediction and ground truth vortex locations were in adjacent pixels.
Metric definitions and justifications for accepting off-by-one errors can be found in \mbox{\ref{sec:validation_metrics}}.

\begin{figure}[t]
    \centering
    \includegraphics[width=1.0\linewidth]{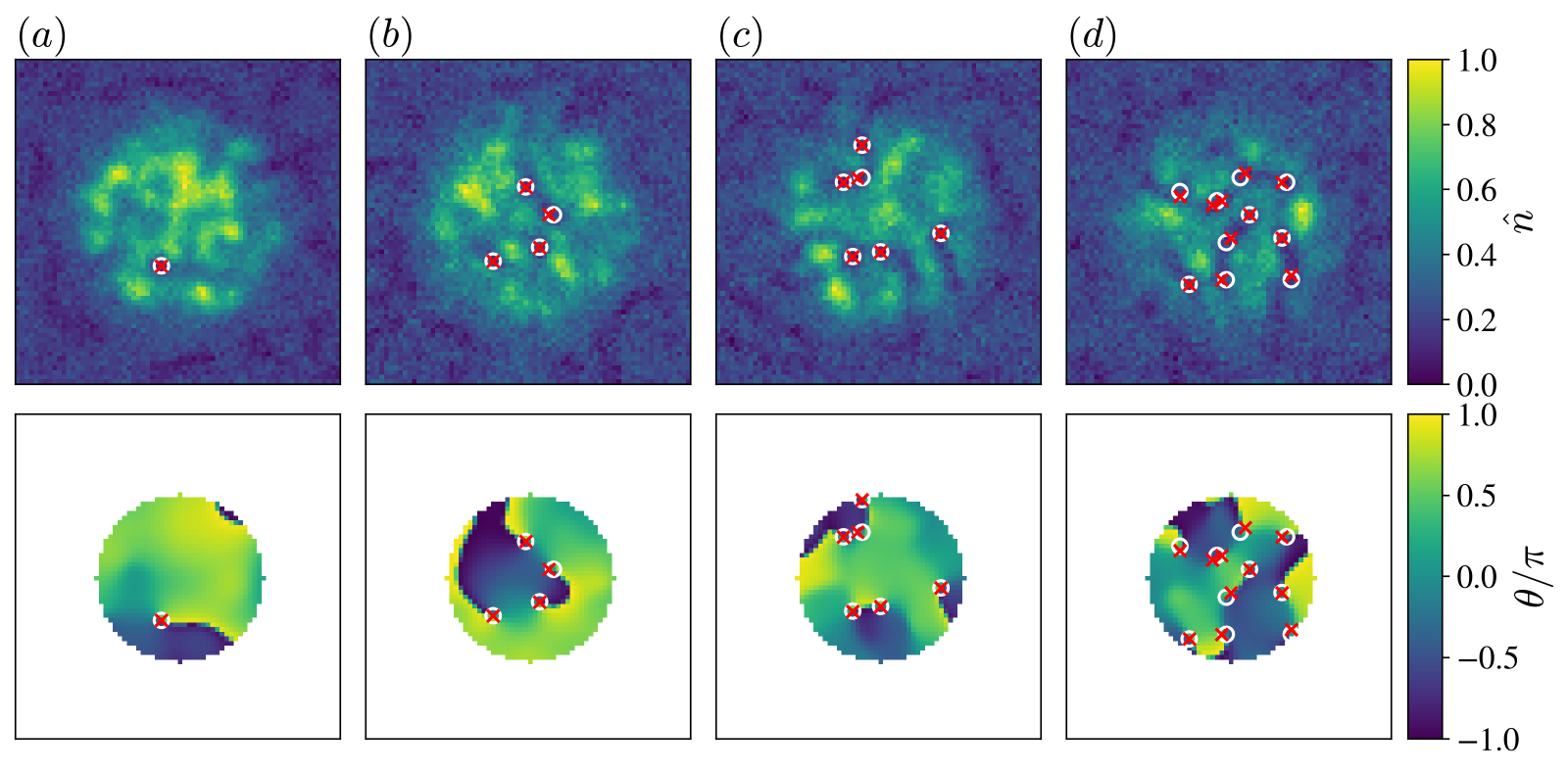}
    \caption{
    Four validation examples (top row), and their corresponding phase profiles (bottom row), with phase-space density decreasing from (a) to (d).
    The predictions (red crosses) and ground-truth vortex locations (white circles) are overlaid in the region of interest, which is defined as the circle with radius equal to 80\% of the cloud radius.
    The phase profiles are only shown for the region of interest to allow easier comparison to the density images.
    The normalized density $\hat{n}$ ranges from 0 to 1 over each image.
    }
    \label{fig:validation_example}
\end{figure}

We found the precision, recall and F1 scores to be 0.81, 0.65 and 0.72, respectively.
We deliberately configured our model to be precision-dominant so we could be confident about any predictions it makes on experimental images.
Examining Figure \ref{fig:validation_example} justifies our leniency with off-by-one errors, and indicates that the model is highly proficient at locating vortices in out-of-sample simulated data.
Assessing the phase profiles, we see that the model can distinguish density dips caused by vortices (phase singularities) from those caused by phonons.
We chose not to validate the model against ordered noise patterns such as stripes or rings since efficient algorithms to remove such noise patterns are readily available \cite{Chen_2011, Song_2020}.

\subsection{Predictions on experimental images}

\begin{figure}[t]
    \centering
    \includegraphics[width=1.0\linewidth]{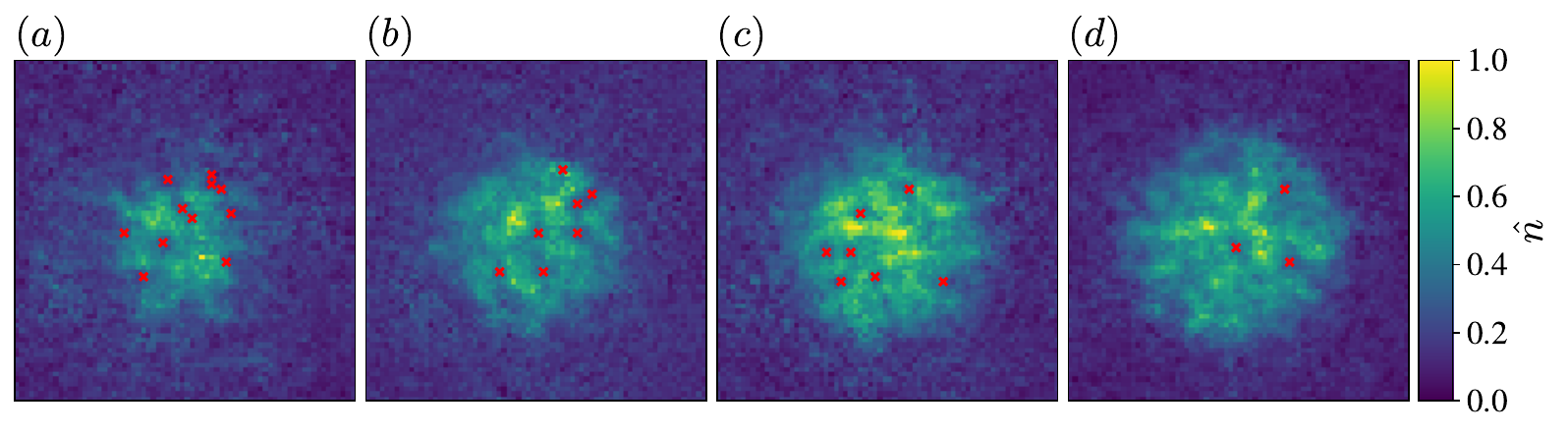}
    \caption{
    Predicted vortex locations for four experimental images.
    The phase-space densities are: (a) 9.3; (b) 10.9; (c) 11.9; (d) 13.5.
    The normalized density $\hat{n}$ ranges from 0 to 1 over each image.
    }
    \label{fig:exp_predictions}
\end{figure}

The experimental dataset had 876 density images, with mean phase-space densities ranging from 8.3 to 16.0, and required preprocessing before we could make vortex predictions.
First, we removed the random variation of the position of the atom cloud by recentering the images on their center of mass.
Then, at each mean phase-space density, we estimated the quasi-condensate radius $r_{\text{QC}}$ by fitting the radial density profile, and defined the region of interest as a circle of radius $0.8r_{\text{QC}}$ as in \cite{Sunami_2022}.
This avoided edge effects where the physics is fundamentally different.
We found that reasonable predictions required keeping the experimental images unnormalized as a result of the low experimental resolution and two fundamental differences in the appearances of vortices that were not captured by the simulation.
Firstly, as phase-space density drops, the fraction of atoms in the non-condensed phase grows which reduces the contrast of density dips.
Secondly, at lower phase-space densities, the reduced detected atom number leads to a lower signal-to-noise ratio.
As such, the signal emanating from potential vortex locations must be amplified in order for the method to work.
The combination of using unnormalized experimental data and applying image contrast enhancement with morphological operators (see \mbox{\ref{sec:data_processing_methods}}) achieved this. The models' predictions are shown in \mbox{Figure \ref{fig:exp_predictions}}.

\begin{figure}[t]
    \centering
    \includegraphics[width=1.0\linewidth]{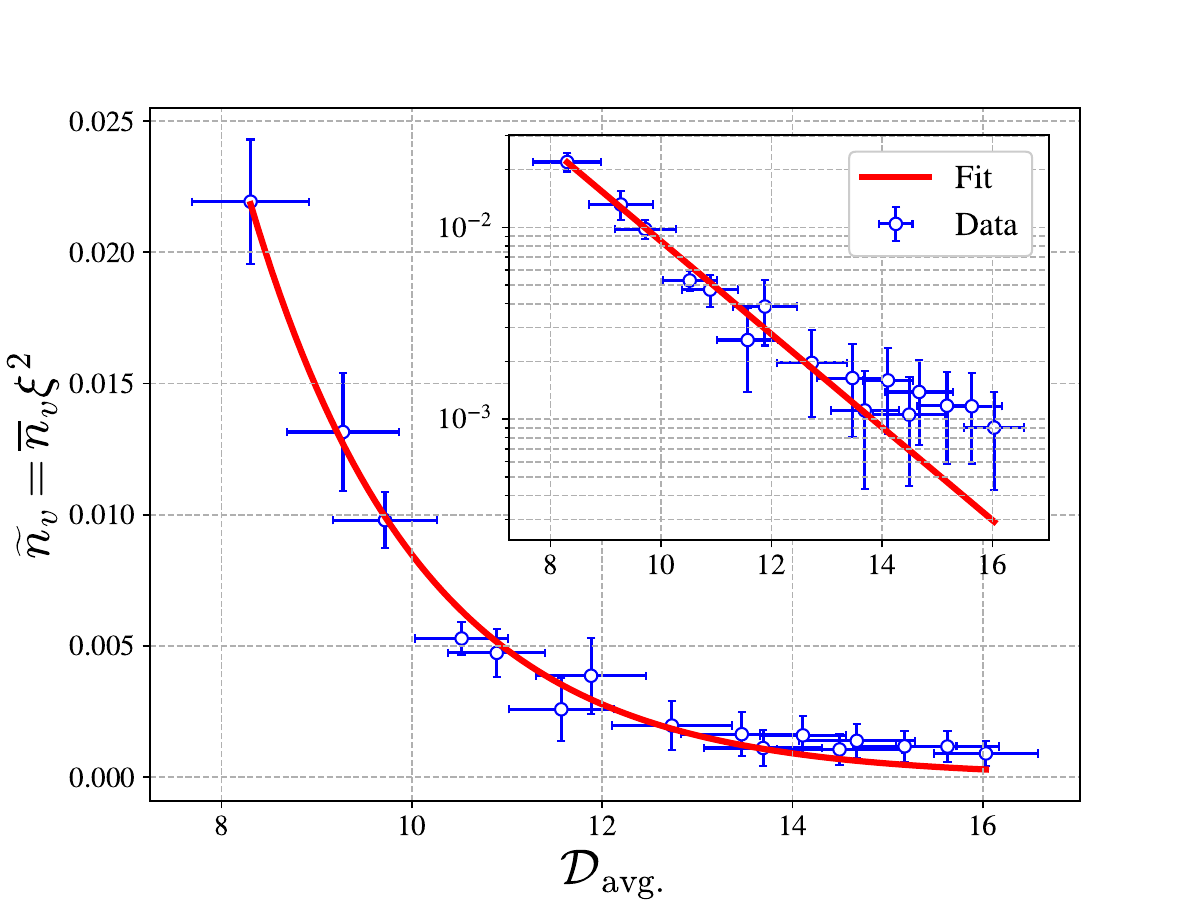}
    \caption{Characteristic scaling of vortex density across the BKT phase transition, from performing inference on experimental images with our CNN; the dimensionless vortex number density $\widetilde{n}_v = \overline{n}_v \xi^2$ is plotted against the mean phase-space density $\mathcal{D}_{\text{avg.}}$. Inset: log-linear plot. The error bars denote the standard error for $\mathcal{D}_{\text{avg.}} < 11$ and Poisson error otherwise.}
    \label{fig:N_vs_psd_exp}
\end{figure}

We could not verify the accuracy of detecting any particular vortex since their positions are not known \textit{a priori} in the experimental images.
However, we analyzed how the dimensionless vortex number density $\widetilde{n}_v = \overline{n}_v \xi^2$, where $\overline{n}_v$ is the mean vortex number density and $\xi$ is the healing length, scales with mean phase-space density $\mathcal{D}_{\text{avg.}}$.
The use of $\widetilde{n}_v$ accounts for the variation of $\xi$ with $\mathcal{D}_{\text{avg.}}$, hence the characteristic area associated with a vortex.
The functional form is expected to follow $\widetilde{n}_v \propto \exp({-b\mathcal{D}_{\text{avg.}}})$ with exponential coefficient \mbox{$b = 0.56(5)$} obtained via matter-wave interferometry \cite{Sunami_2022} and supported by simulations.
Our fit, shown in \mbox{Figure \ref{fig:N_vs_psd_exp}}, yielded \mbox{$b_{\text{exp}} = 0.56(2)$} which is consistent with results using the matter-wave interferometric approach.
The presence of the rapid vortex number increase gives us confidence that we have detected vortices, not phonons, since phonon amplitude grows slowly with decreasing $\mathcal{D}_{\text{avg.}}$ and cannot alone explain the phase transition from a superfluid to a normal fluid.

\section{Summary \& outlook}\label{sec:conclusion}

We have developed a deep learning-based vortex detection method which can infer vortex locations in density images of 2D Bose gases.
The detector is a convolutional neural network (CNN) for which the input is a density image, and the output is a corresponding map of vortex locations.
We used Monte Carlo simulations to obtain a synthetic dataset on which to train, circumventing the time-consuming and error-prone task of labeling vortices by hand, and fulfilling the requirement of a large training set for performant machine learning.
We have shown that our model performs well when evaluated against out-of-sample simulated images scoring 0.81, 0.65 and 0.72 in precision, recall and F1 metrics, respectively.
Further, we have been able to generalize the CNN's predictive power to experimental images where vortices cannot be identified by simple algorithms or the human eye.
These predictions were validated against a standard matter-wave interferometric benchmark from previous work, successfully reproducing the ground-truth vortex number scaling.
Our results demonstrate that the CNN reliably distinguishes vortices from phonons, marking a significant step toward on-demand, system-wide vortex detection in atomic 2D Bose gases.

With improved image quality, the vortices will be more readily detectable, allowing simple and efficient models---such as the one presented in this paper---to make predictions with even greater confidence and accuracy.
Furthermore, transformers can be used to enhance existing CNN-based approaches by integrating global context, better capturing the relationships between vortices, though this comes with increased computational demands.
Looking ahead, applications to real-time tracking of free vortices as described in \cite{Andersen_2006, Navarro_2013, Koukouloyannis_2014}, or vortex pairing dynamics as in \cite{Foster_2010} will become more feasible.
We hope such improvements will allow this approach to extend to bilayer systems such as those in \cite{Sunami_2022, Beregi_2024, Sunami_2024}, where inter-layer interactions are expected to give rise to correlated free vortices in the two layers, which can be distinguished from other vortices by density depletions of different magnitude.

Finally, we anticipate that simulation-supported, deep learning-based vortex detection algorithms, like the one presented in this paper, will be a valuable tool not only in the study of quantum gases, but in programmable quantum simulators in general, a quantum technology experiencing rapid growth in recent years \cite{Schafer_2020, Ebadi_2021}. These algorithms are fast and efficient, and since the underlying simulation can be adjusted to match different scenarios, they are also very adaptable.

\section{Data availability statement}

Example preprocessing, training, and inference notebooks are available at \cite{Sesodia_2024_code} alongside a trained model. Simulated and experimental data is available upon reasonable request.

\section{Acknowledgments}

MS received support from the John O’Connor Research Fund, St Peter's College, Oxford. The data analyzed was from experimental work supported by EPSRC grant EP/X024601/1.
We gratefully acknowledge the support of NVIDIA Corporation with the donation of the Titan Xp GPU used in this research.

\appendix
\section{The BKT transition in 2D Bose gases}\label{sec:bkt_transition}

\begin{figure}[t]
    \centering
    \includegraphics[width=1.0\linewidth]{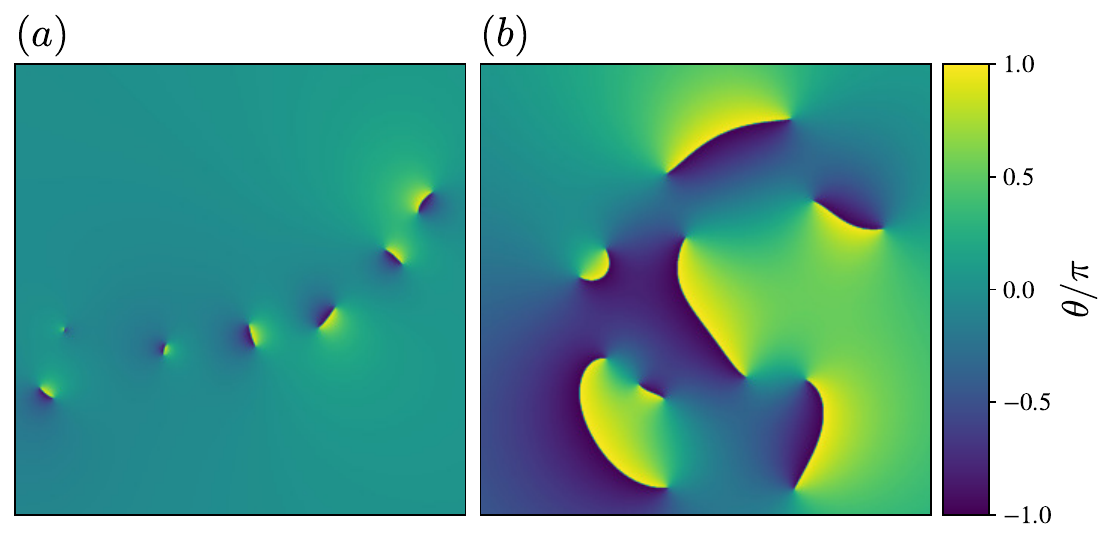}
    \caption{
    Cartoon diagram illustrating the effect of vortex separation on the phase field.
    In (a), seven vortices and seven antivortices are placed such that each vortex is close to an antivortex, that only affect the phase close to the pair.
    In (b), the seven vortices and seven antivortices are placed randomly which scrambles the phase.
    }
    \label{fig:bkt}
\end{figure}

In uniform 2D Bose gases, fluctuations (phonons) destroy true long-range order at any finite temperature, thus inhibiting the formation of a Bose-Einstein condensate, but the system still exhibits superfluid behavior at sufficiently low temperatures.
The phase transition between the superfluid phase at low temperature and the normal phase at high temperature was explained by Berezinskii \cite{Berezinskii_1972}, Kosterlitz, and Thouless \cite{Kosterlitz_1973}, who identified the key microscopic mechanism involving vortices which drives the transition. The transition is between a superfluid, quasi-ordered phase where correlations decay algebraically to a disordered, normal fluid phase where correlations decay exponentially.
At low temperatures, unbound vortices are strongly suppressed due to their large energy cost compared to the thermal energy available, instead closely-bound vortex-antivortex pairs form \cite{Kosterlitz_1974}.
The presence of these closely-bound pairs still allows superfluid behavior as the circulation of the pairs cancels for contours which enclose the pair.
Such vortex pairs only affect the phase significantly on length scales comparable with the pair separation, therefore the local phase of the quantum system remains well-correlated over larger length scales. 
Upon further temperature increase, the vortices in these pairs gain the energy to move further apart until the system reaches the critical temperature at which they unbind, resulting in free vortices moving in the system.
This unbinding scrambles the phase over all relevant length scales, i.e., the quasi-long-range order makes way for disorder, corresponding to a transition from the superfluid phase into the normal phase.
\mbox{Figure \ref{fig:bkt}} illustrates the effect of vortex separation on the phase-field.

\section{Experimental details}\label{sec:exp_details}

We prepare 2D degenerate Bose gases of $^{87}$Rb atoms in a cylindrically-symmetric trap having strong confinement along the $z$ direction.
In this work, we created vertical confinement using the RF dressing techniques described in \cite{Perrin_2017, Luksch_2019, Barker_2020b_jphysb, Barker_2020b_newjphys, Sunami_2022}, with vertical trap frequency of $\omega_z/2\pi = \SI{1.2}{kHz}$. 
This resulted in the dimensionless 2D interaction strength $\widetilde{g} = \sqrt{8\pi} a_s/\ell_0=0.08$, where $a_s$ is the 3D s-wave scattering length, $\ell_0=\sqrt{\hbar/(m\omega_z)}$ is the harmonic oscillator length along $z$ for an atom of mass $m$, and $\hbar$ is the reduced Planck constant. 
We imposed additional optical trapping in the horizontal plane using a ring-shaped, strong, off-resonant laser beam to realize near-homogeneous 2D systems \cite{Sunami_2024}.
We loaded $N$ atoms into the trap at a temperature $T=\SI{50}{\nano \kelvin}$, set by forced evaporation. 
We varied $N$ to cover a broad range of the phase-space density $\mathcal{D} = n h^2 /(2\pi m k_B T)$, where $n$ is the average 2D density.
For all parameters employed, the quasi-2D conditions $\hbar \omega_z > k_B T$ and $\hbar \omega_z > \mu$ were satisfied.

\section{Simulation details \& verification}\label{sec:simulation_verification}

\begin{figure}[t]
    \centering
    \includegraphics[width=1.0\linewidth]{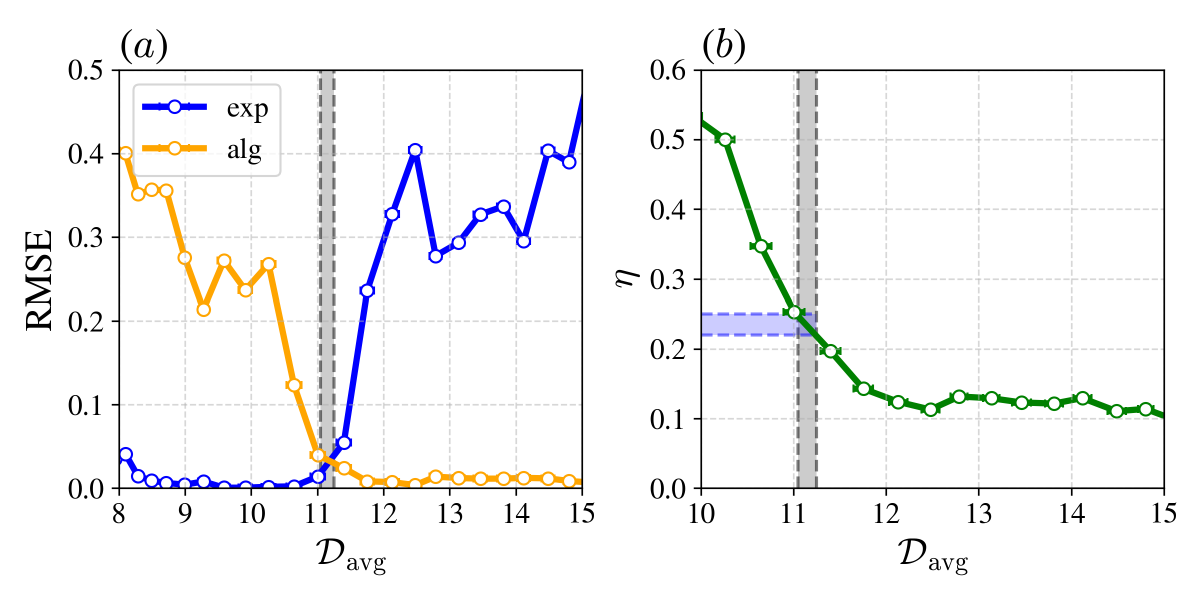}
    \caption{
    Correlation function analysis: (a) The root-mean-square errors of algebraic and exponential fits to the correlation functions are plotted as a function of the mean phase-space density $\mathcal{D}_{\text{avg.}}$; (b) The algebraic exponent $\eta$ is plotted as a function of mean phase-space density $\mathcal{D}_{\text{avg.}}$.
    The gray, vertical shading indicates the crossover region in which the BKT transition occurs.
    }
    \label{fig:corr_analysis}
\end{figure}

To ensure thermal equilibrium and uncorrelated fields, we initially ran the algorithm for 200\,000 updates per site, and then began sampling fields with a further 5000 updates per site occurring between each field sample.
We tuned $\sigma$ such that the mean acceptance rate of field updates was near 0.25, which yields optimal convergence \cite{Gelman_1997, Roberts_2001}.
We collected 2500 fields at eight phase-space densities evenly spaced in the range $\mathcal{D} \in [8.0, 11.2]$ by keeping the temperature constant at $T=50$ nK and varying $\mu$. We expected the phase transition to occur at $\mathcal{D} \approx 11$ for our finite-size system which we verify by direct analysis of the two-point phase correlation function in Figure \ref{fig:corr_analysis}

We can see from Figure \ref{fig:corr_analysis}(a) that the BKT transition occurs around $\mathcal{D} \approx 11$ as evidenced by the narrow crossover region between the root-mean-square errors of algebraic and exponential fits to the correlation function.
In Figure \ref{fig:corr_analysis}(b), we see that the crossover region places the algebraic exponent $\eta$, defined when fitting $r^{-\eta}$ to the correlation function, just below the universal value of $0.25$ found in the thermodynamic limit -- a typical result for finite-sized systems \cite{Sunami_2022}.

\begin{figure}[t]
    \centering
    \includegraphics[width=1.0\linewidth]{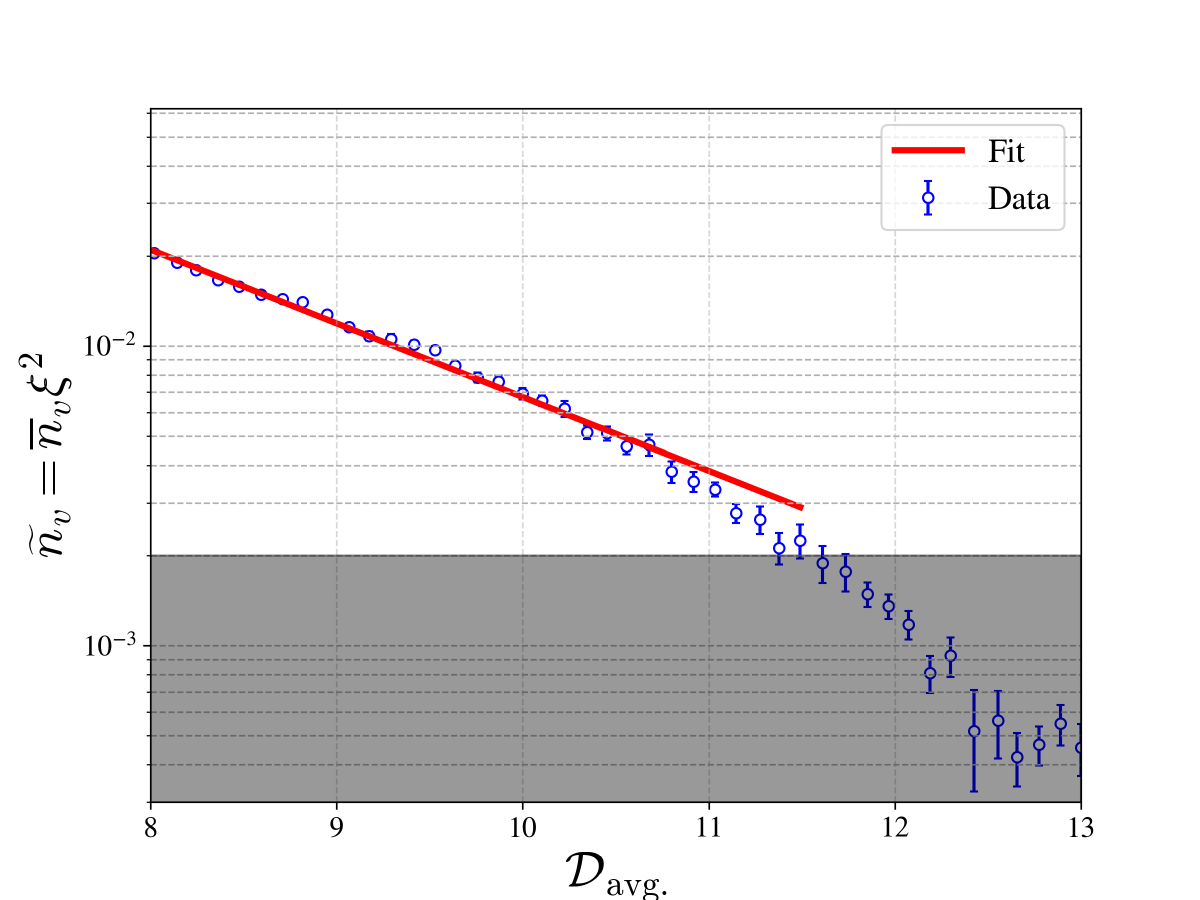}
    \caption{
    Characteristic scaling of vortex density in numerical simulation; the dimensionless vortex number density $\widetilde{n}_v = \overline{n}_v \xi^2$ is plotted as a function of the mean phase-space density $\mathcal{D_{\text{avg.}}}$ on a log-linear scale. 
    For $\mathcal{D} \gtrsim 11$, vortex excitations are sparse and the extraction of number density becomes slightly unstable with finite amount of sampling, with larger uncertainties (gray shaded region).
    }
    \label{fig:char_scaling}
\end{figure}

We verified that our simulation reproduced the scaling of dimensionless vortex number density $\widetilde{n}_v$ with mean phase-space density $\mathcal{D}_{\text{avg.}}$ previously observed in \cite{Sunami_2022}.
Figure \ref{fig:char_scaling} shows the result on a log-linear scale, with our fit region restricted to the range of mean phase-space densities present in our training data ($\mathcal{D}_{\text{avg.}} \in [8.0, 11.2]$).
We found the exponential fit coefficient $b_{\text{sim.}} = 0.57(1)$ to be consistent with previous experimental results.
We ignored the region $\widetilde{n}_v < 2 \times 10^{-3}$ as the vortex counts are extremely low, averaging less than a single vortex per image.

To ensure the lattice accurately represented the continuum limit, we compared the site separation $\delta l = 0.25$ \textmu m to two characteristic length scales for a Bose gas: 1) the healing length $\xi = \hbar/\sqrt{2 m g n}$ which is the characteristic size of the vortex core, and 2) the thermal de Broglie wavelength $\lambda = h/\sqrt{2 \pi k_B T}$.
We found, empirically, that our healing length falls in the range \mbox{$\xi \in [0.59, 0.99]$\,\textmu m}, and at $T = 50$\,nK the thermal de Broglie wavelength is \mbox{$\lambda = 0.85$\,\textmu m}. Thus, \mbox{$\delta l \lesssim \xi, \lambda$}, and our lattice approximation is justified.

\section{Data processing methods}\label{sec:data_processing_methods}

\begin{figure}[t]
    \centering
    \includegraphics[width=1.0\linewidth]{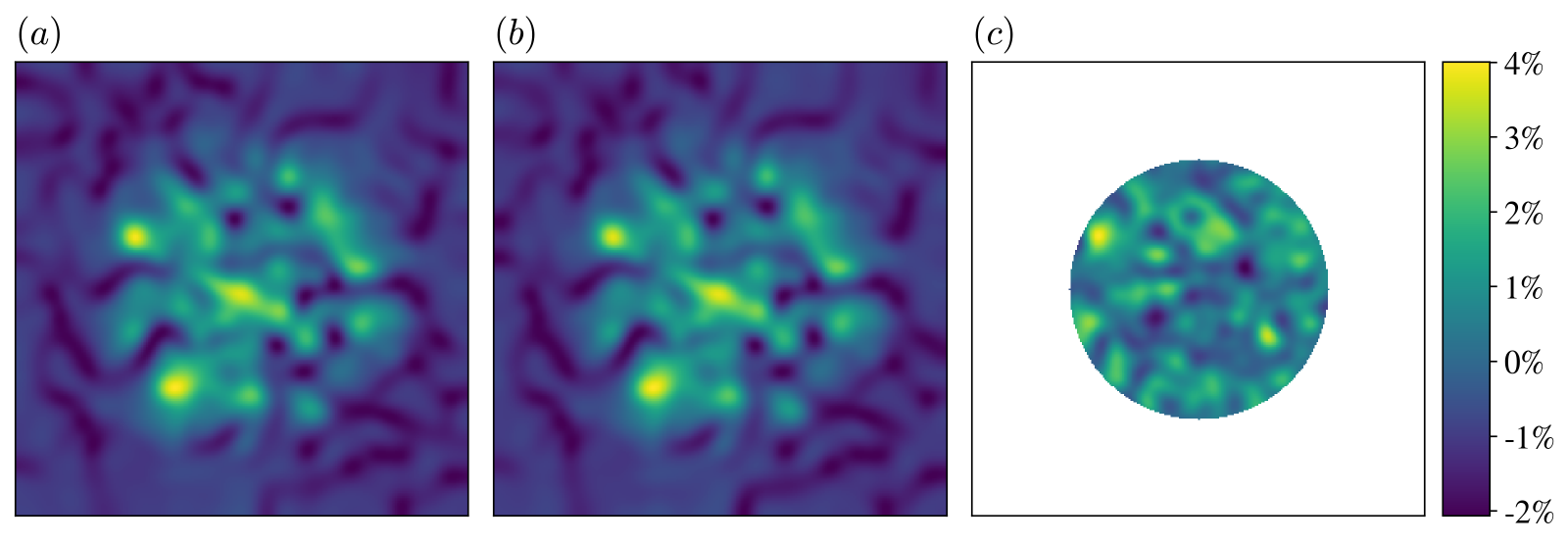}
    \caption{
    Comparison of the densities of a single field using (a) 2D simulation and (b) 3D simulation.
    The colorbar applies to (c) where the relative difference between individual pixel values in (a) and (b) is shown to not be significant.
    }
    \label{fig:2dvs3d}
\end{figure}

We justified the choice to simulate the time-of-flight expansion in 2D, as opposed to in 3D, by plotting the density images that arose in either case. For the 3D simulation, the density was averaged along $z$ in a region of width $10$ \textmu m as in the experiment.
We see, in Figure \ref{fig:2dvs3d}, that the density images are qualitatively indistinguishable, and by subtracting one from the other, we see that the relative difference of any given pixel is at most a few percentage points.

\begin{figure}[t]
    \centering
    \includegraphics[width=1.0\linewidth]{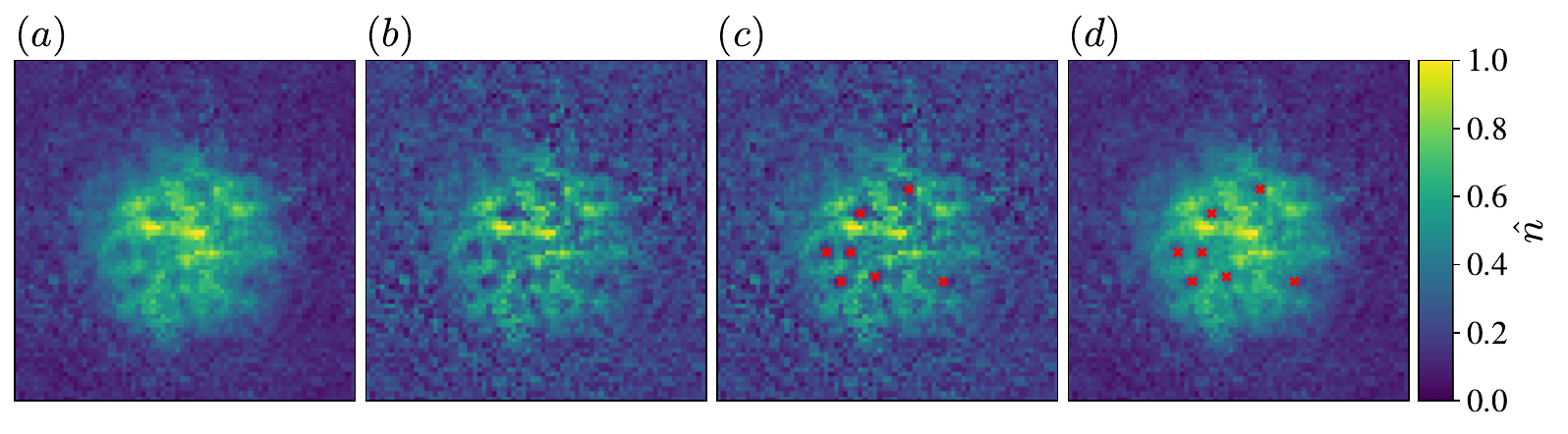}
    \caption{
    Prediction pipeline: The base experimental image (a) has its contrast enhanced resulting in (b).
    Predictions occur on the enhanced image (c) and are then overlayed back onto the original image (d).
    The normalized density $\hat{n}$ ranges from 0 to 1 over each image.
    }
    \label{fig:enhancement_pipeline}
\end{figure}

We applied contrast enhancement, inspired by \cite{Widyantara_2016}, to the experimental images to aid prediction.
In brief, the higher density (bright) and lower density (dark) areas were identified using the morphological operators top-hat and black-hat, respectively.
The bright areas were added to, and the dark areas were subtracted from, the original image to form the enhanced image.
The effect of this enhancement is shown in Figure \ref{fig:enhancement_pipeline}.

\section{Neural network architecture and training details}\label{sec:cnn_details}

Our CNN has four successive convolutional layers of which the first three use the rectified linear unit (ReLU) activation function and the final layer uses sigmoid to obtain a probability between zero and one. 
Each convolutional layer, barring the final one, is followed by a batch normalization layer to stabilize and speed up training \cite{Ioffe_2015}.
The filter sizes and number of filters in each layer are shown in Table \ref{table:nn_architecture}.
Including the batch normalization layers, there are 24\,406 learnable parameters.

Before training, we normalized our combined training and validation dataset to have pixel values in the range [0, 1]. Then, using the loss function hyperparameter $w_1 = 3.3$, a learning rate of $\alpha = 0.0001$, a batch size of 32 and the Adam optimization algorithm \cite{Kingma_2017}, we trained for 10 epochs.
The training PC was equipped with 64\,GB of RAM and a single NVIDIA Titan Xp GPU, resulting in a training time of around 90 seconds, and inference of 1000 images in 350\,ms.

\begin{table}[h]
\begin{indented}
\centering
\item[]\begin{tabular}{@{}ccccr}
\br
Layer & No. Filters & Filter Size & Stride & No. Parameters\\
\mr
    Conv1 & 30 & $5 \times 5$ & 1 & 780 \\
    Conv2 & 25 & $5 \times 5$ & 1 & 18775 \\
    Conv3 & 20 & $3 \times 3$ & 1 & 4520 \\
    Conv4 & 1 & $3 \times 3$ & 1 & 181 \\
\br
\end{tabular}
\end{indented}
\caption{Neural Network Architecture. Batch normalization layers which occur after every convolutional layer are not shown but contain additional parameters.}
\label{table:nn_architecture}
\end{table}

\section{Validation metrics}\label{sec:validation_metrics}

Our evaluation metrics were based on precision and recall which are standard in binary classification. Denoting the number of true positives as TP, false negatives as FN, and similarly for TN and FP, we define precision and recall as

\begin{equation}
    \text{Precision} = \frac{\text{TP}}{\text{TP} + \text{FP}},
\end{equation}

\begin{equation}
    \text{Recall} = \frac{\text{TP}}{\text{TP} + \text{FN}}.
\end{equation}

\noindent The F1 score is defined as the harmonic mean of precision and recall:

\begin{equation}
    \text{F1} = 2 \times \frac{\text{Precision} \times \text{Recall}}{\text{Precision} + \text{Recall}}.
\end{equation}

We allowed for off-by-one errors in the above metrics, and justified this choice as follows.
The characteristic size of an \textit{in situ} vortex is given by its healing length which we found to be in the range \mbox{$\xi \in [0.59, 0.99]$\,\textmu m}.
After time-of-flight (TOF) expansion, the size of the vortex core increases by a factor of two, verified by GPE simulations.
Given our pixel size (1\,\textmu m) was of the same order, we expected individual vortices to have physical extent over multiple pixels.
Additionally, the ground truth locations contained their own ambiguity due to the factor-of-four downsampling.
As such, we expected the model to incur off-by-one errors, where the prediction and ground truth vortex locations are in adjacent pixels, and for the model to sometimes predict the same vortex in adjacent cells.
Since we were only interested in isolated vortices, the off-by-one errors were safely ignored, and the adjacent predictions were fixed by post-processing.

The post-processing was a custom non-max suppression (NMS) algorithm.
NMS is a common post-processing stage in the field of object detection and is typically used to choose a single bounding box for an object when many are generated \cite{Girshick_2014, Redmon_2016}.
Here our NMS algorithm kept only the highest probability vortex location in any given pair of predictions on adjacent cells when both prediction probabilities were greater than 0.5.

\section{Density profiles of detected vortices}\label{sec:vortex_profile}

\begin{figure}[t]
    \centering
    \includegraphics[width=1.0\linewidth]{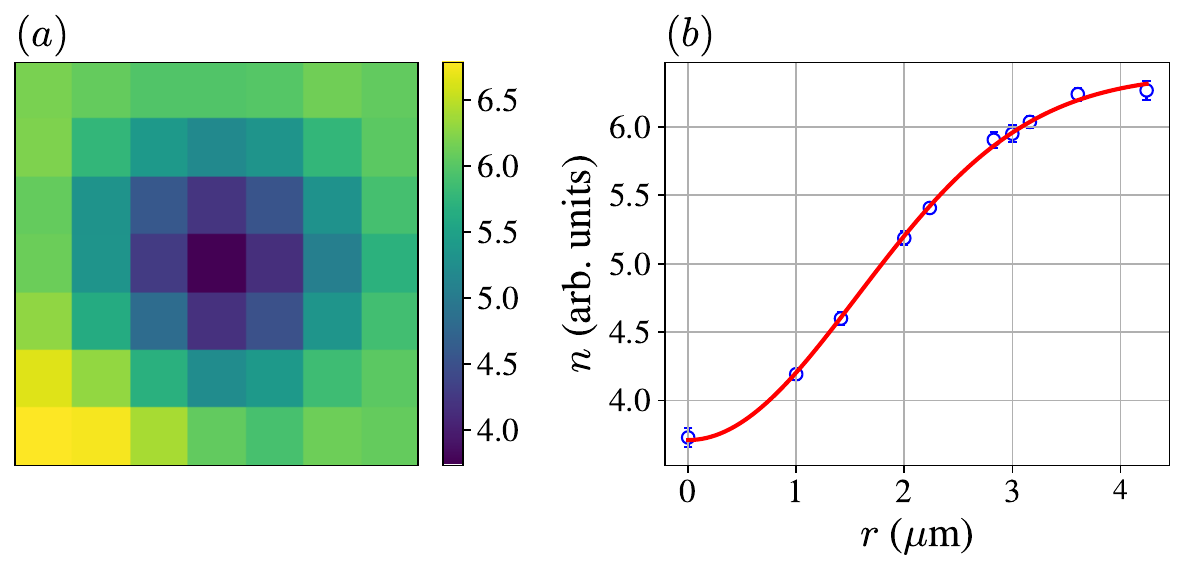}
    \caption{
    Density profiles of vortices detected in experimental data: (a) the average appearance of all the detected vortices at $\mathcal{D} = 11.88$, each pixel is 1\,{\textmu m}$^2$; (b) the radial density profile of (a) fitted with a Gaussian.
    The density $n$ (in arbitrary units) labels both the colorbar in (a) and y-axis in (b).
    }
    \label{fig:vortex_profile}
\end{figure}

To further verify that the CNN had only detected vortices in the experimental images, we calculated the width of the density dips at the predicted vortex locations.
We averaged the $7 \times 7$ regions where vortices were predicted for a given phase-space density (see Figure \ref{fig:vortex_profile}(a)), and fitted a Gaussian to the corresponding radial density profile to estimate the vortex size $r_v$, defined as the half-width at half-maximum of the fit (see Figure \ref{fig:vortex_profile}(b)).
We found $r_v = 1.84(3)$ \textmu m, which is close to the value obtained from zero-temperature GPE simulations of TOF expansion dynamics of a single vortex in a system of comparable density.

\newpage
\section*{References}

\bibliographystyle{iopart-num}
\bibliography{references}

\end{document}